\documentclass[12pt]{article}%
\usepackage{amsfonts}
\usepackage{amssymb}
\usepackage{amsmath}
\usepackage{enumitem}
\usepackage{geometry}
\usepackage{graphicx}
\usepackage{float}

\usepackage{setspace}
\usepackage{multirow}
\usepackage{booktabs}%
\usepackage{natbib}
\providecommand{\U}[1]{\protect\rule{.1in}{.1in}}
\newtheorem{theorem}{Theorem}

\newtheorem{corollary}[theorem]{Corollary}

\newtheorem{result}[theorem]{Result}

\newenvironment{proof}[1][Proof]{\noindent\textbf{#1.} }{\ \rule{0.5em}{0.5em}}
\setlist[enumerate]{leftmargin=*,align=left,label=A\arabic*.}
\newcommand\setItemnumber[1]{\setcounter{enumi}{\numexpr#1-1\relax}}

\begin{document}

\title{ }

\begin{center}
\textit{Original Article}\bigskip

{\LARGE Doubly Robust Regression Analysis for Data Fusion}

{\Large Katherine Evans}$^{\dag}${\Large , BaoLuo Sun}$^{\ddag}${\Large , James Robins}$^{\ast}$, \\
{\Large and Eric J. Tchetgen Tchetgen}$^{\ast\ast}$ 

$^{\dag}${\large Verily Life Sciences LLC}

$^{\ddag}${\large Department of Statistics and Applied Probability, National University of Singapore}

$^{\ast}${\large Departments of Epidemiology and Biostatistics, Harvard T.H. Chan School of Public Health}

$^{\ast\ast}${\large Department of Statistics, The Wharton School of the University of Pennsylvania}

\textbf{Abstract}
\end{center}

\noindent This paper investigates the problem of making inference about a parametric model for the
regression of an outcome variable $Y$ on covariates $(V,L)$ when data are fused from two separate
sources, one which contains information only on $(V, Y)$ while the other contains information
only on covariates. This data fusion setting may be viewed as an extreme form of missing data
in which the probability of observing complete data $(V,L,Y)$ on any given subject is zero. We
have developed a large class of semiparametric estimators, which includes doubly robust
estimators, of the regression coefficients in fused data. The proposed method is DR in that it
is consistent and asymptotically normal if, in addition to the model of interest, we correctly
specify a model for either the data source process under an ignorability assumption, or the
distribution of unobserved covariates. We evaluate the performance of our various estimators via an extensive simulation study, and apply the proposed methods to investigate the relationship between net asset value and total expenditure among U.S. households in 1998, while controlling for potential confounders including income and other demographic variables.


\bigskip\pagebreak

\section{\noindent Introduction}

Parametric likelihood based inference for regression analysis is a well-developed area of modern statistical
theory. In recent years, fairly complete theory has also developed to account for incomplete
outcome or covariate information in regression analysis. Inverse probability weighting (IPW) of complete
cases and multiple imputation are two prominent methods that stand out in modern missing data
theory \citep{robins1994estimation, little2014statistical}. A fundamental assumption on which most missing data methods rely is that the
probability of observing a subject with complete data is bounded away from zero, also known as
the positivity assumption, which is often necessary for identification of the full data law and smooth functionals of the latter \citep{robins1994estimation}. In this paper, we consider a more extreme form of incomplete data,
in which the positivity assumption does not hold, i.e. the probability of observing complete data is
zero for all units in the population.

This situation may arise, for instance, when two data sets from separate sources are fused
together such that no unit belongs to both sources and some variables obtained from one source
are not available in the other source. For instance, as we consider throughout in the paper, it may
be that the outcome of interest $Y$ is collected only in the first data set but not in the second, and
likewise, a subset of regressors $L$ are only observed in the second data set but not in the first. Both
data sets contain information on common variables $V$ . A prominent example of such missing data
structure concerns the main/validation study design in comparative effectiveness studies. In such
design, a main study sample in which outcome, treatment variable and a relatively limited subset
of confounders are available, is enriched with an external validation sample which contains extensive
potential confounders together with treatment information, but lacks outcome information \citep{sturmer2005adjusting}. The
two datasets are then fused together in the hope that information available in the validation sample
can somehow be leveraged to reduce confounding bias.

Another example, somewhat related to meta-analysis for prediction model evaluation \citep{riley2010meta,debray2013framework,debray2017guide},
might involve enriching a data set of a clinical study with covariate information from a separate
source, say a study containing socio-demographic or summary-level information, but no outcome data, for the
purpose of improving clinical risk prediction \citep{chen2000unified,chatterjee2016constrained}. Clearly, in both of these examples, a regression
model for the outcome on the combined set of covariates can be identified only under fairly stringent
parametric assumptions and, as we discuss below, provided that there is a non-trivial overlap in the
amount of information available from both sources of data. We shall refer to this general framework
as regression analysis for data fusion.

The missing data literature has previously described the data fusion problem as that of "statistical
matching". The textbooks by \citet{d2006statistical} and \citet{rassler2012statistical} provide an extensive
overview of the state of the art for data fusion. \citet{d2010old} provides a
comparison of many of the existing data matching methods in the literature and of assumptions
needed to recover valid inferences using these methods. A fundamental assumption on which much
of this literature relies on is that of conditional independence between $Y$ and $L$ given $V$, an assumption
which is likely untenable in practice. This assumption is particularly problematic in the two settings
described above where a potential non-null association between $Y$ and $L$ given $V$ is an important
part of the scientific hypothesis under consideration. When the samples are drawn from a finite
population according to a complex survey design, concatenation \citep{rubin1986statistical} and calibration \citep{renssen1998use, wu2004combining} are two
commonly used methods for statistical matching. Concatenation involves modifying the sample
weights of the samples in order to get a unique sample given by the union of the original sample
with new weights that represent the population of interest. The new weights require computing the
probability of the subjects in one sample under the survey design of the other sample, which requires
detailed knowledge of the survey designs. Calibration preserves both samples and calibrates
the two sets of survey weights. The method obtains a unique estimate of the common variable,
$V$, by combining the estimates of the distribution of $V$ from both samples and then calibrating
the original sample weights to the obtained estimate. The weights are then used to estimate the
distribution $f(L|V)$ in the sample with $L$ and the distribution $f( Y|V)$ in the sample with outcome $Y$. \citet{wu2004combining} suggests similar approaches with different constraints for the sample weights, such as
forbidding negative weights. Recent work by \citet{conti2016statistical} allows estimation of the distribution function of variables not jointly observed in the presence of logical constraints without necessarily imposing the conditional independence assumption, and the corresponding bounds for matching error can be estimated from sample data. \citet{graham2016efficient} introduces a general framework for data combination under moment restrictions and estimators that are doubly robust only under restricted model specification of nuisance parameters.

Data fusion is also prominent in literature on instrumental variable (IV) methods
for causal inference. An instrumental variable is an exogenous variable known to be associated
with a treatment or exposure variable of interest, and to be associated with an outcome of
interest only through its association with treatment. The IV approach can, under certain conditions, be
used to recover an unbiased estimate of a causal effect in the presence of unmeasured confounding.
The most common IV approach assumes a linear model relating the outcome to exposure and
observed covariates, together with a linear model relating exposure to IV and covariates. \citet{angrist1992effect} examine estimation and inference about the causal effect of exposure under
such linear models, when IV and exposure are available from one data source, while outcome and
IV are available in a separate data source, so that no subject has available data on all three
variables, IV, exposure and outcome. These two-sample instrumental variable estimators deliver
point identification and inference by explicitly leveraging parametric assumptions. Regression using
Two-Sample Two-Stage Least Squares was introduced by \citet{klevmarken1982missing} and shown by \citet{inoue2010two} to be more efficient than the two-sample instrumental variable estimator.
These methods assume that both samples are i.i.d. random samples from the same population with
finite fourth moments and are independent. \citet{graham2016efficient} identified the two-sample IV estimation problem as one specific example of a larger, general class of data combination or fusion problems, and derived semiparametric efficiency bounds under the corresponding general class of moment conditions which allow sample moments of the common
variables $V$ to differ significantly across the two datasets being
combined. \citet{pacini2017two} assumes independence of the samples
and makes use of the marginal distributions to provide a characterization of the identified set of the
coefficients of interest when no assumption on the joint distribution of $(Y, V,L)$ is imposed.

\citet{robins1995semiparametric} consider a missing data setting closely related to ours. The main contribution
of their paper is to characterize a large class of semiparametric estimators of a parametric
conditional density of $Y$ given $(L,V)$ when $L$ is missing at random. They characterize in a general semiparametric missing data model with sole restriction a model for the full data, the efficient
influence function for the parameters of the parametric model which is the solution to an integral
equation that is not generally available in closed form. They also point out in a remark that \citet{bickel1993efficient} and \citet{hasminskii1983asymptotic} obtained results similar to theirs when $Y$ and $L$ are never observed together, which is the data fusion setting with which the current paper is
concerned. 

An important contribution of our paper is to show that, in fact, there is a large class of
influence functions for the parameters of the conditional density $f(Y |L, V)$ available in closed form in a missing data model that is otherwise unrestricted, 
and therefore convenient candidates as estimating functions. The proposed semiparametric estimating
functions include doubly robust (DR) estimating functions that yield estimators which are consistent
and asymptotic normal if, in addition to the outcome model of interest, one correctly specifies a
model for either the data source process or the distribution of unobserved covariates. Importantly, unlike \citet{graham2016efficient}, we do not restrict specification of nuisance models to belong to a certain class of models, e.g. their DR result only holds if missing data model is specified as a certain logistic regression model. In addition, we show that the efficient influence function for
the parameters of the conditional density is available in closed form in the special case where the
outcome is polytomous.

In section 2 we lay out notation and assumptions. In section 3 we develop the general class
of estimators as well as a new semi-parametric doubly robust method. In section 4 we discuss
implementation.  In section 5 we discuss
local efficiency in the special case of binary outcome, although the result readily generalizes to polytomous outcome, and provide approximately efficient 
influence functions in the case of continuous outcome. We examine and evaluate the finite sample performance of the double robust
 approach in an extensive simulation study summarized in section 6, and illustrate the proposed methods on fused data from the U.S. Bureau of Labor Statistics' Consumer Expenditure Survey and the Federal Reserve Board's Survey of Consumer Finances in section 7. We conclude in section 8 with
a discussion. Throughout, proofs and derivations can be found in the appendix.

\section{\noindent Notation and Assumptions}

Let $R$ be an indicator that a subject is observed in data source $\mathcal{A}$ $(R = 1)$ or in data source $\mathcal{B}$ $(R = 0)$.
Let $V$ denote covariates which are observed in both sources, $Y$ denote the outcome only
observed in source $\mathcal{A}$, and $L$ denote covariates only observed in source $\mathcal{B}$. The full data $(Y,L, V)$
are i.i.d realizations from a common law $f(Y,L, V )$. Let $f(Y |V,L)$ denote the true conditional
distribution of Y given $(V, L)$. Let $\pi(V)= \Pr(R = 1|V )$ be the probability that a subject is in data
source $\mathcal{A}$. 
Throughout, we make the following assumptions:

\begin{enumerate}
\item Correct outcome model: $f (Y |V,L; \theta)$ is correctly specified such that $f(Y|V,L;\theta^{\dag})=f(Y |V,L)$ for some value $\theta^{\dag}$;
\item Positivity: $\delta < \pi(V) < 1-\delta$ almost surely, for a fixed positive constant $\delta$;
\item Ignorability: $R \perp (Y,L)|V$,
\end{enumerate}
and we let $\mathcal{M}$ denote the set of models which satisfy (A1-3).
Assumption (A1) requires that the outcome model proposed for $f$ is correctly specified.
The positivity assumption (A2) states that the probability of observing a subject in
either data source is bounded away from both 0 and 1. We note that (A2) is strictly weaker
that the usual positivity assumption typically assumed in missing data problems which requires a
positive probability of observing complete data for each subject. Assumption (A3) states that the probability that a unit is observed in either data
source only depends on $V$ and does not further depend on $Y$ or $L$. This assumption is akin to
missing at random and is imposed on the data source process which is technically a nuisance parameter not of primary scientific interest, in contrast to the conditional independence assumption $Y \perp L| V$ imposed on the full data law of primary interest required
by some existing methods, such as matching \citep{d2010old}. 

\section{\noindent IPW and DR Estimating Functions}

In this section we describe a large class of IPW estimating functions for $\theta$ under various sets of modeling assumptions of nuisance parameters. Let
$\pi(V ;\eta) = P(R = 1|V ; \eta)$ denote a parametric model for the data source process indexed by a finite
dimensional parameter $\eta$. We shall make use of the following assumption:

\begin{enumerate}
  \setItemnumber{4}
\item $\pi(V ;\eta)$ is correctly specified such that $\pi(V ;\eta^{\ast})=\pi(V)$ for some value $\eta^{\ast}$.
\end{enumerate}

Let $\mathcal{M}_{\pi}=\mathcal{M} \cap \left\{\pi(V ;\eta) : \eta \right\}$. For user-specified function $g(Y,V)$ of $(Y,V)$, let
\begin{align}
U_g(\theta;\eta)=\frac{R}{\pi(V;\eta)}g(Y,V)-\frac{1-R}{1-\pi(V;\eta)}E_{\theta}[g(Y,V)|V,L]. \label{ef}
\end{align}
Below we discuss assumptions $g(Y,V)$ must satisfy to ensure identification.

\begin{result}
Under  $\mathcal{M}_{\pi}$,
\begin{align}
E_{\eta^{\ast}}\left[  U_g(\theta^{\dag};\eta^{\ast}) \right]=0. \label{ipwef}
\end{align}
\end{result}

The parallel IPW function given in $(\ref{ef})$ assigns to every subject the inverse probability of observing the subject from the data source
in which he or she was indeed observed. Interestingly, this general class of estimating functions includes a large set of DR
estimating functions. Suppose that one has specified a parametric model $t(V;\alpha)$ for the density $t(L|V)$ of $L$ given $V$.
\begin{enumerate}
  \setItemnumber{5}
\item $t(V;\alpha)$ is correctly specified such that $t(V;\alpha^{\ddagger})=t$ for some value $\alpha^{\ddagger}$.
\end{enumerate}

Let $\mathcal{M}_{t}=\mathcal{M} \cap \left\{t(V;\alpha): \alpha \right\}$. Then let
\begin{align}
U^{DR}_g(\theta;\eta,\alpha)=&\frac{R}{\pi(V;\eta)}\left\{g(Y,V)-E_{\theta,\alpha}\left[g(Y,V)|V \right]\right\}\nonumber \\ 
&+\frac{1-R}{1-\pi(V;\eta)}\left\{E_{\theta,\alpha}\left[g(Y,V)|V \right]-E_{\theta}[g(Y,V)|V,L]\right\}. \label{dr}
\end{align}

\begin{result}
Under the union model  $\mathcal{M}_{\pi \cup t}=\mathcal{M}_{\pi} \cup \mathcal{M}_{t}$,
\begin{align}
E_{\eta^{\ast},\alpha^{\ddagger} }\left[  U^{DR}_g(\theta^{\dag};\eta, \alpha) \right]=0, \label{dref}
\end{align}
if either $\eta=\eta^{\ast}$ or $\alpha=\alpha^{\ddagger}$, but not necessarily both. 
\end{result}

Estimating function (\ref{dr}) is said to be doubly robust for $\theta$ in that estimators based on (\ref{dr}) are consistent for $\theta^{\dagger}$
provided we correctly specify a model for $t (V; \alpha)$ or $\pi(V ; \eta)$, but
not necessarily both. Additionally, when both models are correctly specified, the estimator for $\theta$ based on $U^{DR}_g(\theta;\eta, \alpha)$ is most efficient (for a fixed choice of $g$) in $\mathcal{M}_{\pi \cup t}$.

We note that due to the DR property of the estimating function given in (\ref{dr}), its unbiasedness still holds for any choice of $\pi(V)$, if the conditional density $t(V;\alpha)$ is correctly specified. Heuristically the resulting estimator works by correctly imputing the missing values in $L$ conditional on $V$. For user-specified function $g(Y,V)$, let
\begin{align}
U^{imp}_g(\theta;\alpha)&=U^{DR}_g(\theta^{\dag};\pi=0.5 , \alpha) \\
&\propto {R}\left\{g(Y,V)-E_{\theta,\alpha}\left[g(Y,V)|V \right]\right\}+{(1-R)}\left\{E_{\theta,\alpha}\left[g(Y,V)|V \right]-E_{\theta}[g(Y,V)|V,L]\right\}. \label{imp}
\end{align}

\begin{corollary}
Under  $\mathcal{M}_{t}$,
\begin{align}
E_{\alpha^{\ddagger}}\left[  U^{imp}_g(\theta^{\dag}; \alpha^{\ddagger}) \right]=0. \label{impef}
\end{align}
\end{corollary}

In the next section, we construct feasible IPW, imputation (IMP) and DR estimators as solutions to empirical versions of (\ref{ipwef}), (\ref{dref}) and (\ref{impef}) respectively, and describe the large sample behavior of the resulting estimators of $\theta$.

\section{\noindent IPW, IMP and DR Estimation}
Feasible IPW, IMP and DR estimators involves a first-stage estimation of nuisance parameters $\eta$ and $\alpha$. 
We propose the following estimator of $\eta$ which maxmizes the log-likelihood,
\begin{align}
\hat{\eta}=\text{arg max} \sum_i \left\{ R_i \log \pi(V_i;\eta)+(1-R_i)\log[1-\pi(V_i;\eta)] \right\} \label{alpha}
\end{align}

By ignorability assumption (A3), $\alpha$ can be estimated by likelihood maximization
restricted to sample $\mathbb{B}$. That is,
\begin{align}
\hat{\alpha}=\text{arg max} \left\{\sum_i (1-R_i)\log t(L_i|V_i; \alpha)\right\}.
\end{align}

Let $\mathbb{P}_n$ denote the empirical mean operator $\mathbb{P}_n f(O) = n^{-1}\sum_i f(O_i)$. Then the IPW, IMP and DR estimates of $\theta$ are solutions to the estimating functions $\mathbb{P}_n \left\{ U_g(\theta;\hat{\eta}) \right\}=0$, $\mathbb{P}_n \left\{ U^{imp}_g(\theta;\hat{\alpha}) \right\}=0$ and $\mathbb{P}_n \left\{ U^{DR}_g(\theta;\hat{\eta},\hat{\alpha}) \right\}=0$ respectively.
Under standard regularity conditions given in Theorem 2.6 of \citet{newey1994large}, the resulting IPW estimator of $\theta$ is consistent if $\pi(V;\eta)$ is correctly specified and the DR estimator is consistent if either $\pi(V;\eta)$  or $t(V; \alpha)$, but not necessarily both, is correctly specified. 

 To illustrate, suppose that we have univariate $Y$, $p$-dimensional $ L$, and $q$-dimensional $V$ which are all continuous, with a constant term embedded in $V$.  Let $A^T$ denote the transpose of $A$. IPW estimation proceeds by first obtaining $\hat{\eta}$. For example, assuming a logistic model $\pi(V;\eta) =\left(1+\exp^{- V^T \eta}\right)^{-1}$,  we then solve (\ref{alpha}) by fitting logistic regression on observed data $(R,V)$. DR estimation additionally requires the estimate $\hat{\alpha}$. Suppose the conditional density of $L$ given $V$ is multivariate normal $ \mathcal{N}(\alpha^T V, \Sigma)$, where the errors in $\Sigma$ may be correlated but do not vary among observations. The $q \times p$ estimate $\hat{\alpha}$ can be computed via least squares estimation, $\hat{\alpha}=\left(V_{\mathbb{B}}^T V_{\mathbb{B}} \right)^{-1}V_{\mathbb{B}}^T L_{\mathbb{B}}$, where $\left(V_{\mathbb{B}}, L_{\mathbb{B}}\right)$ is the $n \times (p+q)$ covariate matrix from data source $\mathbb{B}$ with $n$ observations. Finally, we assume that $Y|V,L$ is normally distributed as $ \mathcal{N}(\beta^T (V^T,L^T)^T, \Sigma)$, $\theta=\left(\beta, \Sigma \right)$.  If we are primarily interested in the mean parameters $\beta$ and not the variance component $\Sigma$, then a convenient choice for $g(Y,V)$ is given by $Y g(V)$ where $g(V)$ is of the same dimension as $\beta$, and we have the following set of estimating functions:
\begin{align}
U_g(\theta;\eta)&=g(V)\left\{ \frac{R}{\pi(V;\eta)}Y-\frac{1-R}{1-\pi(V;\eta)}E_{\theta}[Y|V,L]\right\}, \label{ipwexample}\\
U^{DR}_g(\theta;\eta,\alpha)&=g(V)\left\{\frac{R}{\pi(V;\eta)}\left\{Y-E_{\theta,\alpha}\left[Y|V \right]\right\} \right. \nonumber\\ 
&\left. \phantom{-}+\frac{1-R}{1-\pi(V;\eta)}\left\{E_{\theta,\alpha}\left[Y|V \right]-E_{\theta}[Y|V,L]\right\}\right\} \label{drexample} \\
U^{imp}_g(\theta;\alpha)&=g(V)\left\{{R}\left\{Y-E_{\theta,\alpha}\left[Y|V \right]\right\} +{(1-R)}\left\{E_{\theta,\alpha}\left[Y|V \right]-E_{\theta}[Y|V,L]\right\}\right\}, \label{impexample}
\end{align}
where $E_{\theta}[Y|V,L]=\beta^T (V^T,L^T)^T$ and $E_{\theta,\alpha}[Y|V]=\beta^T (V^T,V^T \alpha)^T$.
 
In general, if we are interested in estimating the full set of parameters $\theta$ which indexes the assumed parametric model $f(Y|V,L;\theta)$, the choice of $g(Y, V )$ should be such that it is of at least the same dimension as $\theta$, $E\left[U_g^T(\theta)U_g(\theta)\right]<\infty$ and $E\left[\frac{\partial}{\partial \theta}U_g(\theta,\eta) \right]$ is nonsingular. We note that the generalized method of moments (GMM) approach can be adopted to obtain estimates if the choosen function $g(Y, V )$ is of larger dimension than $\theta$.

Let $\phi$ denote the set of nuisance parameters, i.e. $\phi=\eta$, $\phi=(\eta,\alpha)$ and $\phi=\alpha$ for IPW, DR and imputation estimation respectively, and let $\phi^{\ast}$ denote the probability limit of $\hat{\phi}$. The scores for nuisance parameters are
\begin{align*}
S_\eta &= \frac{d}{d \eta} \log \left\{\pi(V;\eta)^R\left[1-\pi(V;\eta) \right]^{1-R} \right\}\\
S_\alpha &= \frac{d}{d \alpha} \log \left\{ t(L|V; \alpha)^{1-R}\right\}.
\end{align*}
Let $S_{\phi}=S_{\eta}$, $S_{\phi}=\left(S^T_\eta, S^T_\alpha\right )^T$ or $S_{\phi}=S_{\alpha}$ for IPW, DR and imputation-based estimation respectively, and let
\begin{equation*}
U_{\theta,\phi}=
\begin{cases}
      \left(U^T_g(\theta;\eta), S^T_{\phi} \right)^T, & \text{for IPW} \\
      \left(U^{DR,T}_g(\theta;\eta,\alpha), S^T_{\phi}\right )^T, & \text{for DR estimation} \\
      \left(U^{imp,T}_g(\theta;\alpha), S^T_{\phi}\right )^T, & \text{for imputation-based estimation}.
    \end{cases}
\end{equation*}
In addition, let 
\begin{align*}
G_{\theta}&=E\left[\frac{\partial}{\partial \theta} U_{\theta^{\dagger},\phi^{\ast}}\right]\\
G_{\phi}&=E\left[\frac{\partial}{\partial \phi} U_{\theta^{\dagger},\phi^{\ast}}\right]\\
M &= E\left[\frac{\partial}{\partial \phi}  S_{\phi^{\ast}} \right] \\
\Psi &= -M^{-1}S_{\phi^{\ast}},
\end{align*}
where all the expectations are evaluated at the true parameter values. Then under standard regularity conditions given in Theorem 6.1 of \citet{newey1994large}, 
\begin{align}
\sqrt{n} \left(\hat{\theta}-\theta ^{\dagger}\right)\xrightarrow[]{d}\mathcal{N}(0,\Sigma_{\theta}),
\end{align}
where
\begin{align}
\Sigma_{\theta}= G^{-1}_{\theta} E\left\{\left[ U_{\theta^{\dagger},\phi^{\ast}}+G_{\phi}\Psi\right]  \left[ U_{\theta^{\dagger},\phi^{\ast}}+G_{\phi}\Psi\right] ^T\right\} G^{-1, T}_{\theta}. \label{var}
\end{align}
For inference, a consistent estimate $\hat{\Sigma}_{\theta}$  of the asymptotic covariance matrix given in (\ref{var}) can be constructed by replacing all expected values with empirical averages evaluated at $\left(\hat{\theta}, \hat{\phi} \right)$. Then a 95\% Wald confidence interval for $\theta_j$ is
found by calculating $\hat{\theta}_j \pm 1.96 \hat{\sigma}_j$, where $\hat{\sigma}_j$ is the square root of the $j^{th}$ component of the diagonal of $n^{-1}\hat{\Sigma}_{\theta}$. Alternatively, nonparametric bootstrap can be performed to obtain estimates of the variance.

\section{\noindent Local Efficiency}
For binary $Y$, any function $g(\cdot)$ of $Y$ and $V$ can be expressed as $g(Y,V)=Yg_1(V)+g_0(V)$, where $g_1(\cdot)$ and $g_0(\cdot)$ are arbitrary functions of $V$. Therefore the class of DR estimating functions in (\ref{dr}) is equivalently given by
$$
\mathcal{L}_{DR}= \left\{g_1(V) M(\theta): g_1(\cdot) \text{ arbitrary}   \right\},
$$
where 
$$
M(\theta)=\frac{R}{\pi(V;\eta)}\left\{Y-E_{\theta,\alpha}\left[Y|V \right]\right\}+\frac{1-R}{1-\pi(V;\eta)}\left\{E_{\theta,\alpha}\left[Y|V \right]-E_{\theta}[Y|V,L]\right\}.
$$
We have the following result:
\begin{result}
Suppose $\hat{\theta}_h$ is a regular and asymptotically linear (RAL) estimator of $\theta$ in the semiparametric model $\mathcal{M}_{\pi \cup t}$. Then, 
\begin{align*}
\sqrt{n}\left(\hat{\theta}_h-\theta^{\dag}\right)\xrightarrow[]{D} \mathcal{N} \left(0,E\left[h(V)\nabla_{\theta}M(\theta) \right]^{-1} E\left\{M^2(\theta^{\dag})  h(V)h(V)^T\right\} E\left[h(V)\nabla_{\theta}M(\theta) \right]^{-1T}\right)
\end{align*}
for some $h(V) M(\theta)\in \mathcal{L}_{DR}$. $\hat{\theta}_{\hat{h}}$ achieves the semiparametric efficiency bound for $\mathcal{M}_{\pi \cup t}$ at the intersection submodel $\mathcal{M}_{\pi} \cap \mathcal{M}_{t}$ if $\hat{h}$ converges in probability to
\begin{align*}
h^{opt}(V)=-E\left[\nabla_{\theta}M(\theta)|V \right]E\left[M^2(\theta)|V \right]^{-1}.
\end{align*}
\end{result}

Result 3 can easily be extended to polytomous $Y$ with $s>2$ levels using a similar approach by noting that $g(Y,V) = \sum_{k=1}^{s-1} I(Y=y_k)g_k(V)+g_0(V)$ and therefore 
$$
\mathcal{L}^s_{DR}= \left\{\sum^{s-1}_{k=1}g_k(V) M_k(\theta): g_k(\cdot) \text{ arbitrary for }k=1,2,...,s-1   \right\},
$$
where 
\begin{align*}
M_k(\theta)=&\frac{R}{\pi(V;\eta)}\left\{I(Y=y_k)-P(Y=y_k|V; \theta,\alpha)\right\}\\
&+\frac{1-R}{1-\pi(V;\eta)}\left\{P(Y=y_k|V; \theta,\alpha)-P(Y=y_k|V,L; \theta)\right\}, \phantom{-}k=1,2,...,s-1.
\end{align*}
When $Y$ contains continuous components, the semiparametric efficient influence function for $\theta$ is in general not available in closed form, in
the sense that it cannot be explicitly expressed as functions of the true
distribution \citep{robins1995semiparametric}. Let $L_2\equiv L_2(F)$ denote the Hilbert space of zero-mean functions of $p$ dimensions, $Z\equiv z(V,Y)$, with inner product $E_{F}\left(Z^T_1Z_2 \right)=E\left(Z^T_1Z_2 \right)$, and the corresponding squared norm $||Z||^2=E\left(Z^T_1Z_2 \right)$, where $F$ is the distribution function that generated the data. We adopt the general strategy proposed in \citet{NEWEY1993419} (see also \citet{tchetgen2009doubly}) to obtain an approximately locally efficient estimator by taking a basis system $\psi_j(Y, V)$ $(j = 1,...)$ of functions dense in $L_2$, such as tensor products of trigonometric, wavelets or polynomial bases for controls $V$ and $Y$. For approximate efficiency, in practice we let the $p$-dimensional $g_K(Y,V)=\tau \Psi_K$ where $\tau \in \mathbb{R}^{p\times K}$ is a constant matrix and $\Psi_K=\left\{ \psi_1, \psi_2,...,\psi_K   \right\}^T$ for some finite $K>p$. 

To derive an approximately locally efficient estimator for $\theta$, let $\mathcal{K}$ denote the linear operator
$$
\mathcal{K}(\cdot) = \frac{R}{\pi(V;\eta)}\left\{\cdot-E_{\theta,\alpha}\left[\cdot|V \right]\right\}+\frac{1-R}{1-\pi(V;\eta)}\left\{E_{\theta,\alpha}\left[\cdot|V \right]-E_{\theta}[\cdot|V,L]\right\},
$$ 
defined over the space of arbitrary functions of $Y$ and $V$ in $L_2$. Consider the class of influence functions of the form
$$
\mathcal{L}_{\Psi_K}= \left\{\tau\mathcal{K}(\Psi_K)=\tau\left[\mathcal{K}(\psi_1),\mathcal{K}(\psi_2),...,\mathcal{K}(\psi_K)  \right]^T: \tau  \in  \mathbb{R}^{p\times K}  \right\}.
$$
Analogous to Result 3, it can be shown based on Theorem 5.3 in \citet{newey1994large} that the efficient estimator of all estimators with influence functions of the form in $\mathcal{L}_{\Psi_K}$ is indexed by the constant matrix
$$
\tau^{opt}=-E\left[\nabla_{\theta}\mathcal{K}(\Psi_K)\right]E\left[\mathcal{K}(\Psi)\mathcal{K}^T(\Psi_K)\right]^{-1}.
$$
 In particular, the inverse of the asymptotic variance of the estimator indexed by $\tau^{opt}$ is 
\begin{align*}
\Omega_K &= E\left\{ \nabla _{\theta}\mathcal{K}(\Psi_K)\right\}^T {E}\left\{\mathcal{K}(\Psi_K)\mathcal{K}^T(\Psi_K)\right\}^{-1}E\left\{ \nabla _{\theta}\mathcal{K}(\Psi_K) \right\}\\
&=E\left\{ S_{\theta}\mathcal{K}^T(\Psi_K) \right\} {E}\left\{\mathcal{K}(\Psi_K)\mathcal{K}^T(\Psi_K)\right\}^{-1}E\left\{S_{\theta}\mathcal{K}^T(\Psi_K)\right\}^T,
\end{align*} 
evaluated at $\theta=\theta^{\dag}$, and $S_{\theta}$ is the score vector with respect to $\theta$. Thus, $\Omega_K$ is the variance of the population least squares regression of $S_{\theta}$ on the linear span of $\mathcal{K}(\Psi_K)$. Since $\Psi_K$ is dense in $L_2$, as the dimension $K \to \infty$ the linear span of $\mathcal{K}(\Psi_K)$ recovers the subspace in the orthocomplement nuisance tangent space $\Lambda^{\perp}$ containing the efficient score $S_{\theta, \text{eff}}$ so that $\Omega_K  \to || \Pi\left(S_{\theta}|\Lambda^{\perp}\right)||^2=\text{var}\left(S_{\theta, \text{eff}}\right)$, the semiparametric information bound for estimating $\theta^{\dag}$ in the union model $\mathcal{M}_{\pi \cup t}$.

\section{\noindent Simulation Study}
In this section, we report a simulation study evaluating the finite sample performance of our proposed
estimators involving i.i.d. realizations of $(R,RY,(1-R)L,V)$. For each of the sample sizes $n=500, 2000$, we simulated 1000 datasets as followed:
\begin{align*}
C&\sim \mathcal{N}(0, 0.5^2), \phantom{-}A|C \sim \mathcal{N}\left(\lambda_0 + \lambda_1C,  \sigma_A^2\right), \phantom{-}V = (A,C)\\
L|V&\sim \mathcal{N} (\alpha_0+\alpha_1A+\alpha_2C+\alpha_3AC, \sigma^2_L)\\
R|V &\sim \text{Bernoulli}\left\{\pi(V;\eta)\right\},\phantom{-} \pi(V;\eta)=\left(1+\exp^{-\eta_0-\eta_1A+\eta_2C} \right)^{-1}\\
Y|V,L &\sim \mathcal{N}(\beta_0+\beta_1A +\beta_2 C + \beta_3 L, \sigma^2_Y),
\end{align*}
with $(\lambda_0,\lambda_1,\sigma_A) = (0.5, 0.5, 0.3)$, $(\alpha_0,\alpha_1,\alpha_2,\alpha_3,\sigma_L)=(-0.5,1.5, 1.0, 2.0, 0.3)$, $(\beta_0,\beta_1,\beta_2,\beta_3,\sigma_Y)=(0.5,-0.5,1.0,1.5, 0.4)$  and $(\eta_0,\eta_1,\eta_2)=(0.5,-0.75,-0.75)$  so that marginally $\Pr(R=1)\approx 0.5$. Our aim is to estimate the conditional mean parameters $\beta=(\beta_0,\beta_1,\beta_2,\beta_3)$ based on the observed data, by solving empirical versions of (\ref{ipwexample}-\ref{impexample}) for IPW, DR and imputation-based estimation respectively with $g(V)=\left(1,A,C,AC\right)^T$ using the R package ``BB"  \citep{varadhan2009bb}. In each simulated sample, we estimated the proposed estimators' asymptotic variance given by (\ref{var}), and Wald 95\% confidence interval coverage
rates were computed across the 1000 simulations.

We also evaluated the performance of the proposed estimators in situations where
some models may be mis-specified. Let superscript $\mathsection$ denote probability limits from fitting the misspecified models.
The data source model was misspecified as $\tilde{\pi}$ by dropping $C$ from the logistic model,
that is, $\tilde{\pi}(V;\eta^{\mathsection})=\left(1+\exp^{-\eta^{\mathsection}_0-\eta^{\mathsection}_1A} \right)^{-1}$. The density of $L|V$ was misspecified as $\tilde{t}$ by fitting a standard linear regression using only $(C,C^2)$ as regressors, i.e. $E[L|V; \alpha^{\mathsection}]=\alpha^{\mathsection}_0+\alpha^{\mathsection}_1C+\alpha^{\mathsection}_2C^2$. We explored four scenarios corresponding to (i) correct models $\pi$ and $t$, (ii) correct $t$ but incorrect model $\tilde{\pi}$, (iii) correct $\pi$ but incorrect model $\tilde{t}$ and (iv) incorrect models $\tilde{\pi}$ and $\tilde{t}$. Figure \ref{fig:1} present results for estimation of the regression coefficient $\beta_3$, while Table \ref{tab:1} shows the corresponding empirical coverage rates; the results for the remaining regression coefficients $(\beta_0,\beta_1,\beta_2)$ are qualitatively similar and therefore relegated to the appendix.

\begin{figure}[t!]
   \centering
    \caption{{Boxplots of inverse probability weighted (IPW),  imputation-based (IMP) and doubly-robust (DR) estimators of the regression coefficient $\beta_3$, whose true value of 1.5 is marked by the horizontal line, when $\alpha_3=2$.}}
       \includegraphics[page=1,width=0.8\textwidth]{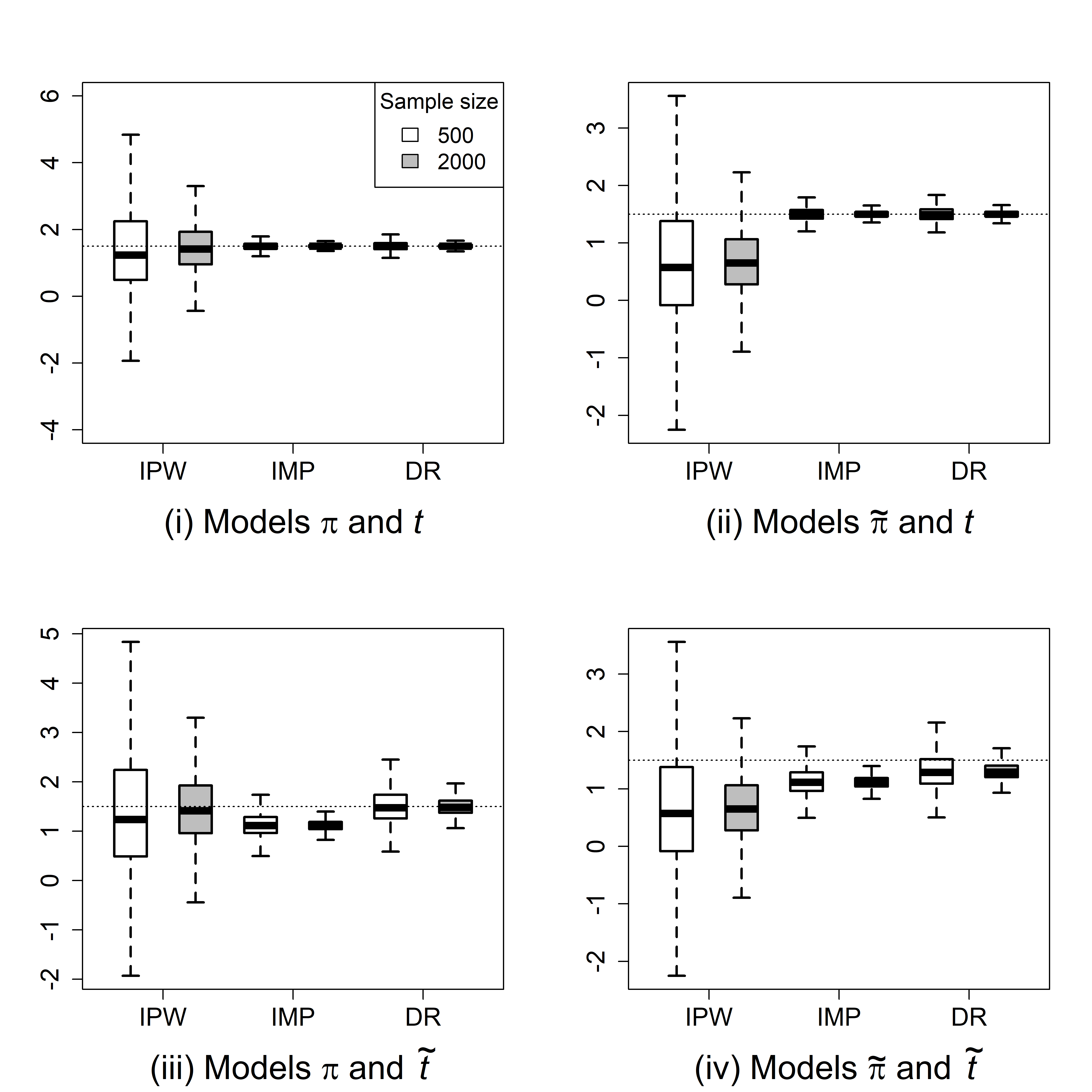} 
        \label{fig:1}
\end{figure}

\begin{table}[t!]
\caption{{Empirical coverage rates based on 95\% Wald confidence intervals, as well as accuracy of standard deviation estimator, under four scenarios: (i) correct $\pi$ and $t$, (ii) correct $t$ but incorrect $\tilde{\pi}$, (iii) correct $\pi$ but incorrect $\tilde{t}$ and (iv) incorrect $\tilde{\pi}$ and $\tilde{t}$. In each scenario, the first row presents results for $n=500$ and the second row for $n=2000$.}}
\label{tab:1}\par
\vskip .2cm
\centering
\renewcommand{\arraystretch}{0.7}
\begin{tabular}{c c c c c c c c c c}
\hline\noalign{\smallskip}
& \multicolumn{3}{c}{Coverage} & \multicolumn{3}{c}{SD ratio$^{\dagger}$} \\
\cmidrule(lr){2-4}\cmidrule(lr){5-7}
&IPW &  IMP & DR &IPW &  IMP & DR  \\
\hline\noalign{\smallskip}
\multirow{2}{*}{(i)} & 0.916 & 0.935 &0.926 &1.178 & 0.955 & 0.899\\ 
& 0.948 & 0.939 & 0.938 & 1.141 & 0.972 &  0.958\\
\multirow{2}{*}{(ii)} & 0.801 & 0.935 &0.923 & 1.164 & 0.955 &0.913 \\ 
& 0.681 & 0.939 &  0.941 & 1.125 & 0.972 & 0.958\\
\multirow{2}{*}{(iii)} & 0.916 & 0.553 & 0.888& 1.178 & 0.939 &0.876  \\ 
& 0.948 & 0.139 & 0.938 & 1.141 & 1.038 & 0.998 \\
\multirow{2}{*}{(iv)} & 0.801 & 0.553& 0.740 & 1.164 & 0.939 & 0.894 \\ 
&0.681 & 0.139 & 0.634 & 1.125 & 1.038 & 1.016 \\
\hline
  \noalign{\vspace{6pt}}%
\multicolumn{5}{l}{$^{\dagger}:$\footnotesize{ Estimated SD / Monte Carlo SD}}
\end{tabular}
\end{table}

 Under correct model specifications (i), the IPW estimator has a small bias at $n=500$ which diminishes  with increasing sample size, while the DR and imputation-based estimators have negligible bias. In agreement with our theoretical results, the IPW estimator is significantly biased in scenarios (ii) and (iv) where the data source process is incorrectly modeled as $\tilde{\pi}$, while the DR estimator shows negligible bias across the scenarios (i)-(iii) and only exhibits significant bias in scenario (iv) where both models are mis-specified as $\tilde{\pi}$ and $\tilde{t}$. The imputation-based estimator shows little bias in scenarios (i) and (ii), but exhibits significant bias in scenarios (iii) and (iv) with misspecified $\tilde{t}$. Under the data generating mechanism considered in this simulation study, the imputation-based estimator is more efficient than the DR estimator, which is in turn more efficient than the IPW estimator across all the scenarios considered. The efficiency of the DR estimator is reduced to a greater extent by mis-specification of $t$ rather than $\pi$. In scenarios where the IPW, DR and imputation-based estimators are unbiased, empirical coverage rates are slightly lower than 0.95 at $n=500$, but approaches the nominal rate with increasing sample size.

For the second set of simulations, we reduce the coefficient for the interaction between $A$ and $C$ in the model for generating $L$ by setting $\alpha_3=0.5$, with all other parameters unchanged.  We require that $E\left[\frac{\partial}{\partial \theta}U_g(\theta,\eta) \right]$ be nonsingular and therefore under the data generating mechanism of this simulation study $L$ and $V$ need to be correlated. We lowered the level of interaction in order to show
how the strength of the relationship between $L$ and $V$ can affect estimation. When the effect of $(A,C)$ interaction in the model that generates $L$ is weak, using $AC$ in $g(V)$ leads to increase in finite-sample bias for all the estimators, and efficiency decreases as well, as shown in Figure \ref{fig:2}.

\begin{figure}[t!]
   \centering
    \caption{{Boxplots of inverse probability weighted (IPW),  imputation-based (IMP) and doubly-robust (DR) estimators of the regression coefficient $\beta_3$, whose true value of 1.5 is marked by the horizontal line, when $\alpha_3=0.5$.}}
       \includegraphics[page=1,width=0.8\textwidth]{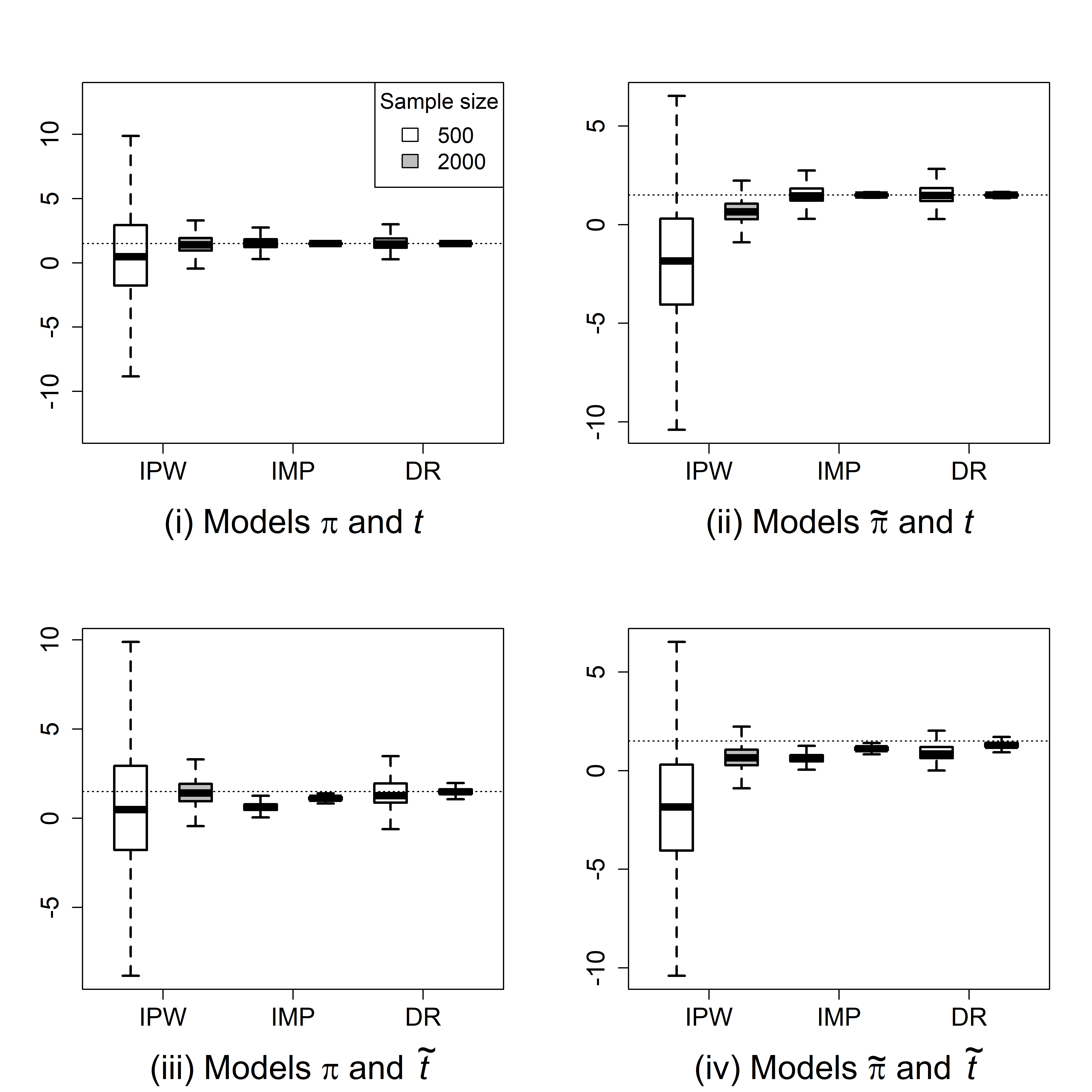} 
        \label{fig:2}
\end{figure}

\section{\noindent Application}

As an empirical illustration, we apply the proposed methods to investigate the relationship between asset value ($L$) and consumption ($Y$), while controlling for potential confounders including income and other demographic variables ($V$). Previous research by \citet{bostic2009housing} leverages on fused data from the U.S. Bureau of Labor Statistics' Consumer Expenditure Survey (CEX) which contains detailed U.S. household expenditure information $Y$, and the Federal Reserve Board's Survey of Consumer Finances (SCF) which provides detailed information on household assets and liabilities $L$, housing and other demographic characteristics. For this application the model of substantive interest is $E(Y|V,L)=(V^T,L)\beta$, and we perform the proposed IPW and DR estimation for $\beta$ based on household expenditure and net worth data from CEX's 1997 fourth quarter survey and 1998 SCF respectively, along with demographic information that is recorded in both surveys. The variables considered in this analysis are presented in Table \ref{tab:2}.

\begin{table}[t!]
\caption{{U.S. household (HH) variables used in the analysis.}}
\label{tab:2}\par
\vskip .2cm
\centering
\renewcommand{\arraystretch}{0.7}
\begin{tabular}{lllllll}
\hline\noalign{\smallskip}
&Variable & Description \\
\hline\noalign{\smallskip}
R & &  Data source indicator for CEX ($R=1$) or SCF ($R=0$) \\
\noalign{\smallskip}
Y &log(expd)& Log of total HH expenditures in fourth quarter of 1997 \\
\noalign{\smallskip}
L & log(netw)& Log of HH total net worth in 1997\\
\hline\noalign{\smallskip}
\multirow{8}{*}{V} & sex &  Sex of HH head (male=0, female=1)  \\
\noalign{\smallskip}
& age & Age of HH head \\
\noalign{\smallskip}
& single &Marital status of HH head (married=0, single=1)  \\
\noalign{\smallskip}
& edu1 & HH head with high school diploma or GED (no=0, yes=1) \\
\noalign{\smallskip}
& edu2 & HH head with some college or Associate degree (no=0, yes=1) \\
\noalign{\smallskip}
& edu3 & HH head with Bachelors degree or higher (no=0, yes=1) \\
\noalign{\smallskip}
& white& White HH head  (no=0, yes=1) \\
\noalign{\smallskip}
& black& Black/African American HH head (no=0, yes=1) \\
\noalign{\smallskip}
&log(income)& Log of total HH income before taxes in 1997\\
\hline
\end{tabular}
\end{table}

 While the data source process is large administrative, the 1998 SCF oversamples relatively wealthy families based on an index created by grossing up capital income flows
observed in the tax data \citep{kennickell1998list}. For IPW estimation the data source model $\pi(V)$ is specified as a logistic regression with main effects for binary variables and up to quadratic terms for age and log(income). In particular, total household income before taxes in 1997 is included in $V$ which may serve as a good proxy for the wealth index in the SCF's sampling design. For DR estimation, we additionally specify $E[L|V]$ as a linear model involving main effects for binary variables and up to quadratic terms for age and log(income) in $V$. We solve empirical versions of (\ref{ipwexample}-\ref{impexample}) for IPW, DR and imputation-based estimation respectively with $g(V)$ specified as a vector which includes the main effects of variables in $V$, and additionally the variable 
$\text{log(income)}^2$ as a correlate for household net worth which is only recorded in the SCF.  We restrict the sample to household heads between 25 and 65 years of
age to mitigate heterogenous consumption effects during college-age years and retirement, and truncate the SCF sample at $90^{th}$ percentiles of observed total household income and net worth due to oversampling of wealthy households in the SCF \citep{bostic2009housing}. The final data set consists of $n=5919$ households (3388 from CEX and 2531 from SCF) for analysis. Due to missing values in the original survey data, the publicly available microdata from both CEX and SCF consists of five imputed replicates; estimation is performed for each replicate and the pooled results using Rubin's rule \citep{rubin2004multiple} are presented in Table \ref{tab:3}.

The DR and imputation-based standard errors are smaller than those from IPW, in agreement with theoretical and simulation results. IPW results suggest that households with married heads generally have greater total expenditures, holding the remaining variables at fixed values. Higher levels of education for the household head is also progressively associated with greater total expenditures.  Finally, after controlling for income and other demographic variables, results from IPW suggest there is a negative association between household net worth and total expenditure, although this is not statistically significant at $0.05$-level. The results from DR and imputation-based estimation generally agree with each other, and statistically significant relationships include an inverse association between age and total expenditure, as well as a positive association between household net worth and total expenditure. We note that both these associations agree qualitatively with the findings from \citet{bostic2009housing}. The similarity between DR and imputation estimates suggests that the conditional model $E[L|V]$ may be specified nearly correctly \citep{robi}, and \citet{tchetgen2010semiparametric} describe a formal specification test to detect which of the two baseline models $\pi(V)$ and $t(V)$ is correct under the union model $\mathcal{M}_{\pi \cup t}$. Based on this and the DR property, it may be that the data source model in this illustrative analysis for IPW is misspecified, and the results from DR estimation are more meaningful given its additional protection against misspecifications of the data source model.

\phantom{$^\ast$}
\begin{table}[t!]
\caption{{Estimates of conditional mean parameters $\beta$ for log total household expenditure. Pooled standard errors are given in brackets, and asterisks denote significance at $0.05$-level.}}
\label{tab:3}\par
\vskip .2cm
\centering
\renewcommand{\arraystretch}{0.7}
\begin{tabular}{ccccccc}
\hline\noalign{\smallskip}
Variable & IPW & IMP & DR \\
\hline\noalign{\smallskip}
 sex &  \phantom{-}2.338$^\ast$ (0.284) & \phantom{-}0.048\phantom{$^\ast$} (0.067) & \phantom{-}0.030\phantom{$^\ast$} (0.058)\\
\noalign{\smallskip}
 age & \phantom{-}0.399 \phantom{$^\ast$}(0.247) & -0.264{$^\ast$} (0.054) & {-}0.160$^\ast$ (0.042) \\
\noalign{\smallskip}
single & -4.109$^\ast$ (0.367) & \phantom{-}0.048\phantom{$^\ast$} (0.055) & \phantom{-}0.023\phantom{$^\ast$} (0.042)  \\
\noalign{\smallskip}
 edu1 & \phantom{-}0.491\phantom{$^\ast$} (0.254) & \phantom{-}0.016\phantom{$^\ast$} (0.081) & \phantom{-}0.083\phantom{$^\ast$} (0.079)\\
\noalign{\smallskip}
edu2 & \phantom{-}0.886{$^\ast$} (0.358)& \phantom{-}0.038\phantom{$^\ast$} (0.094)& \phantom{-}0.081\phantom{$^\ast$} (0.098) \\
\noalign{\smallskip}
 edu3 &\phantom{-}1.373$^\ast$ (0.460)&-0.001\phantom{$^\ast$} (0.113)& \phantom{-}0.035\phantom{$^\ast$} (0.123) \\
\noalign{\smallskip}
 white& \phantom{-}0.580$^\ast$ (0.229) & -0.094\phantom{$^\ast$} (0.086)&-0.052\phantom{$^\ast$} (0.083) \\
\noalign{\smallskip}
black& \phantom{-}0.237\phantom{$^\ast$} (0.269) & \phantom{-}0.134\phantom{$^\ast$} (0.096)& \phantom{-}0.002\phantom{$^\ast$} (0.104)\\
\noalign{\smallskip}
log(income)& \phantom{-}0.537\phantom{$^\ast$} (0.432) & -0.095\phantom{$^\ast$} (0.096)& \phantom{-}0.085\phantom{$^\ast$} (0.066) \\
\noalign{\smallskip}
 log(netw)& -0.620\phantom{$^\ast$} (0.417)& \phantom{-}0.499$^\ast$ (0.089)& \phantom{-}0.346$^\ast$ (0.066)\\
\hline
\end{tabular}
\end{table}

\section{\noindent Discussion}

Traditional regression models break down when two data sources are fused together such that no
subject has complete data. Investigators often consider parametric models for a given outcome
regressed on a number of independent variables, but current parametric models do not adequately
deal with the missing data structure that arises from data fusion. In this paper we have developed
a general class of semiparametric parallel inverse probability weighting estimating functions, whose
resulting estimators are consistent if the outcome regression and data source process are correctly
specified. This general class of estimating functions includes a large set of doubly robust estimating
functions which additionally require a model for the covariates that are missing. An estimator in this class  is DR in that it is consistent and asymptotic normal if we correctly specify
a model for either the data source process or the distribution of unobserved covariates, but not
necessarily both.

There are several areas for additional research on this topic, notably the open question
of how to generalize this method to other settings. A clear extension is the setting of fusing multiple datasets together, not just two. Consider $m$ data sources with $V$ observed
for all and each of $(L_1,L_2,...,L_{m-1},Y)$ observed in only one source with respective indicators of observation $(R_1,R_2,...,R_{m-1},R_m)$
and inclusion probabilities $(\pi_1,\pi_2,...,\pi_{m-1},\pi_m)$. Therefore the observed data are $O=(V,R_1L_1,R_2L_2,...,R_{m-1}L_{m-1},R_m Y)$.
Then, for example, it is easy to extend (\ref{ipwexample}) for linear models to be 
\begin{align*}
U^m_g (\beta)=g(V)\left\{\frac{R_m}{\pi_m}Y-\left[\beta_0 +\frac{R_1}{\pi_1}\beta^T_1 L_1+\frac{R_2}{\pi_2}\beta^T_2 L_2+...+\frac{R_{m-1}}{\pi_{m-1}}\beta^T_{m-1} L_{m-1}+\beta^T_m V \right] \right\},
\end{align*}
 provided $V$ is rich enough for identification.

\bibliographystyle{agsm}
\bibliography{refs} 

\section*{\noindent Appendix}

\subsection*{Derivation of DR linear space}

The observed data likelihood is given by
$$L(O) = f(R|V;\eta)\left\{\int f(Y|V,L; \theta)dF(L|V;\alpha)  \right\}^R f(L|V;\alpha)^{1-R}   f(V;\epsilon),$$
where we consider $\alpha$ and $\epsilon$ to be possibly infinite-dimensional nuisance parameters and  $O=(R,RY,(1-R)L,V)$. The nuisance tangent space is $\Lambda_{\eta} \oplus \Lambda_{\alpha} \oplus \Lambda_{\epsilon} $, where
\begin{align*}
\Lambda_{\epsilon}&=\left\{B_1 S_{\epsilon}(V): E[S_{\epsilon}(V)]=0  \right\}\\
\Lambda_{\alpha}&=\left\{B_2E[S_{\alpha}(V,L)|O]=B_2 \left\{RE[S_{\alpha}(V,L)|Y,V]+(1-R)S_{\alpha}(V,L)\right\}: E[S_{\alpha}(V,L)|V]=0  \right\} \\
\Lambda_{\eta}&=\left\{B_3\left[\frac{\partial}{\partial \eta} \log f(R|V;\eta) \right]  \right\}.
\end{align*}
Let $\Lambda^{\perp}$ be the observed-data linear space that is orthogonal to $\Lambda_{\epsilon}\oplus\Lambda_{\alpha}$. Then for given $h(O) \in \Lambda^{\perp}_{\epsilon,\alpha}$ we have
\begin{align*}
E\left[h(O)S_{\epsilon}(V)\right]&=0 \phantom{-}\forall S_{\epsilon}(V) \in \Lambda_{\epsilon},\\
E\left\{h(O)E[S_{\alpha}(V,L)|O]\right\}&=E\left\{E[h(O)S_{\alpha}(V,L)|O]\right\}\\
&=E\left\{h(O)S_{\alpha}(V,L)\right\}=0\phantom{-}\forall S_{\alpha}(V,L).
\end{align*}
From the results of \citet{robins1995semiparametric} and \citet{hasminskii1983asymptotic}, $\Lambda^{\perp}_{\epsilon,\alpha}$ is given by
\begin{align*}
\Lambda^{\perp}_{\epsilon,\alpha}&=\left\{ Bh(O): E[h(R,V)|V]=0 \text{ or } E[h(O)|L,V]=0\right\}\\
&=\left\{ B\left[ \frac{R}{\pi(V)}\left[g(Y,V)+k(V)\right]-\frac{1-R}{1-\pi(V)}E[g(Y,V)+k(V)|V,L] \right]: g,k\text{ arbitrary, }g(0,x)=0\right\}.
\end{align*}
Therefore, when the data source process is modeled, a typical element in the ortho-complement $\Lambda^{\perp}$ to the nuisance tangent space is given by
$$
\left\{h(O)-\Pi\left[h(O)| \Lambda_{\eta}\right] : h(O)\in  \Lambda^{\perp}_{\epsilon,\alpha}\right\},
$$
where $\Pi$ denotes the projection operator. For a fixed choice of function $g(Y,V)$, the space of elements in $\Lambda^{\perp}$ is a translation of a linear space away from the origin. Specifically, this linear space is given by $V(g)=x_0+M$, with the element
$$
x_0= \left\{\frac{R}{\pi(V)}g(Y,V)-\frac{1-R}{1-\pi(V)}E[g(Y,V)|V,L]\right\}-\Pi\left[\{\cdot\}| \Lambda_{\eta}  \right]
$$
and linear subspace
$$
M= \left\{\left[\frac{R}{\pi(V)}-\frac{1-R}{1-\pi(V)}\right]k(V)\right\}-\Pi\left[\{\cdot\}| \Lambda_{\eta}  \right]=\Pi[ \Omega(V) | \Lambda_{\eta}^{\perp} ].
$$
It is clear that $\Lambda_{\eta}\subset \Omega(V)$. By Theorem 10.1 of \citep{tsiatis2007semiparametric}, the optimal influence function (in terms of smallest variance) for fixed $g(Y,V)$ is given by 
$$
\mathbb{IF}^*(g)= \left\{\frac{R}{\pi(V)}g(Y,V)-\frac{1-R}{1-\pi(V)}E[g(Y,V)|V,L]\right\}-\Pi\left[\{\cdot\}| \Omega(V)\right].
$$

Let 
$$
\left[\frac{R}{\pi(V)}-\frac{1-R}{1-\pi(V)}\right]k^0(V) \in \Omega(V)
$$
be the projection $\Pi\left[\{\cdot\}| \Omega(V)\right]$. Then $k^0(V)$ needs to satisfy 
\begin{align*}
&E\left\{ \left\{\frac{R}{\pi(V)}\left[g(Y,V)-k^0(V)\right]-\frac{1-R}{1-\pi(V)}\left[k^0(V)-E[g(Y,V)|V,L]\right] \right\} \left\{ \left[\frac{R}{\pi(V)}-\frac{1-R}{1-\pi(V)}\right]k(V)\right\} \right\}\\
&=E\left\{ k(V)\left\{ \frac{1}{\pi(V)}\left[E[g(Y,V)|V]-k^0(V)\right]+\frac{1}{1-\pi(V)} \left[k^0(V)-E[g(Y,V)|V]\right] \right\}  \right\} =0\phantom{-}\forall k(V).
\end{align*} 
By assumption (A2), since $\delta < \pi(V) < 1-\delta$ almost surely, $k^0(V)=E[g(Y,V)|V]$ and the DR linear space is given by
$$
\mathcal{L}_{DR}=\left\{\mathbb{IF}^*(g): g(Y,V) \text{ arbitrary}   \right\},
$$
where
$$
\mathbb{IF}^*(g)= \left\{\frac{R}{\pi(V)}\left[g(Y,V)-E[g(Y,V)|V]\right]+\frac{1-R}{1-\pi(V)}\left[E[g(Y,V)|V]-E[g(Y,V)|V,L]\right]\right\}.
$$

\newpage
In the following, expectations are evaluated at the true parameter values.

\begin{proof}[Proof of Result 1]
\begin{align*}
E_{\eta, \theta}\left\{U_g(\theta;\eta) \biggr\rvert V,L \right\}=&E_{\eta, \theta}\left\{\frac{R}{\pi(V)}g(Y,V)-\frac{1-R}{1-\pi(V)}E_{\theta}[g(Y,V)|V,L]\biggr\rvert V,L\right\}\\
=&E_{\theta}[g(Y,V)|V,L]-E_{\theta}[g(Y,V)|V,L] =0.
\end{align*}
\end{proof}

\begin{proof}[Proof of Result 2 (DR property)]
\\
{\bf \noindent Case 1: $\pi(V)$ is correct but $\tilde{t}(L|V)$ is incorrect} \\
Unbiasedness of DR estimating function follows from Result 1 by taking $g^{\prime}(V,L)=g(V,L)+k(V)$; the proof does not involve $\tilde{t}(L|V)$.

{\bf \noindent Case 2: $\tilde{\pi}(V)$ is incorrect but ${t}(L|V)$ is correct} \\
\begin{align*}
E_{\theta, \eta,\alpha}\left\{U^{DR}_g(\theta; \eta,\alpha) \biggr\rvert V\right\}=&E_{\theta, \eta,\alpha}\left\{\frac{R}{\tilde{\pi}(V)}\left\{g(Y,V)-E_{\theta,\alpha}[g(Y,V)|V]\right\}\right.\\
&\left.+\frac{1-R}{1-\tilde{\pi}(V)}\left\{E_{\theta,\alpha}[g(Y,V)|V]-E_{\theta}[g(Y,V)|V,L]\right\}\biggr\rvert V\right\}\\
=&\frac{\pi(V)}{\tilde{\pi}(V)} \left\{E_{\theta,\alpha}[g(Y,V)|V]-E_{\theta,\alpha}[g(Y,V)|V]\right\}\\
&+\frac{1-\pi(V)}{1-\tilde{\pi}(V)}\left\{E_{\theta,\alpha}[g(Y,V)|V]-E_{\theta,\alpha}[g(Y,V)|V]\right\}=0.
\end{align*}
\end{proof}

\begin{proof}[Proof of Result 3]

The proof is based on the following lemma which is part of Theorem 5.3 in \citet{newey1994large}.

{\bf \noindent Lemma S1.}

If $\exists\tilde{h}(V)$ satisfying
\begin{align*}
-E\left[h(V)\nabla_{\theta}M(\theta) \right] =  E\left[M^2(\theta)  h(V)\tilde{h}(V)^T\right] \phantom{-}\forall  h(V), 
\end{align*}
then the estimator indexed by $\tilde{h}(V)$ is most efficient.

\begin{proof}[Proof of Lemma S1]

If $h(V)$ and $\tilde{h}(V)$  satisfy the equality in lemma S1 then the difference of the asymptotic variances of the respective estimators indexed by them is as follows:
\begin{align*}
 &E\left[M^2(\theta)  h(V)\tilde{h}(V)^T\right]^{-1} E\left[M^2(\theta)  h(V){h}(V)^T\right]E\left[M^2(\theta)  \tilde{h}(V){h}(V)^T\right]^{-1}-E\left[M^2(\theta)  \tilde{h}(V)\tilde{h}(V)^T\right]^{-1} \\
=&E\left[M^2(\theta)  h(V)\tilde{h}(V)^T\right]^{-1} E\left[UU^T \right]E\left[M^2(\theta)  \tilde{h}(V){h}(V)^T\right]^{-1},
\end{align*}
where $U=h(V)-E\left[M^2(\theta)  h(V)\tilde{h}(V)^T\right] E\left[M^2(\theta)  \tilde{h}(V)\tilde{h}(V)^T\right]^{-1}\tilde{h}(V)$ and $E\left[UU^T \right]$ is positive semi-definite.
\end{proof}

We show that if $\tilde{h}(V)$ satisfies the equality in lemma S1 then $\tilde{h}(V)=h^{opt}(V)$. 
\begin{align*}
&-E\left[h(V)\nabla_{\theta}M(\theta) \right] =  E\left[M^2(\theta)  h(V)h^{opt}(V)^T\right] \phantom{-}\forall  h(V), \\
\iff & E\left\{ h(V)\left[ M^2(\theta)  h^{opt}(V)  +  \nabla_{\theta}M(\theta) \right]^T \right\}=0  \phantom{-}\forall  h(V), \\
\iff  &E\left\{ h(V)E\left[ M^2(\theta)  h^{opt}(V) +  \nabla_{\theta}M(\theta) \biggr\rvert V\right]^T \right\}=0  \phantom{-}\forall  h(V), \\
\implies &E\left\{ E\left[ M^2(\theta)  h^{opt}(V)  +  \nabla_{\theta}M(\theta) \biggr\rvert V\right]^{\otimes2} \right\}=0 ,\\
\implies &E\left[ M^2(\theta)  h^{opt}(V)  +  \nabla_{\theta}M(\theta) \biggr\rvert V\right] =0 ,\\
\iff &h^{opt}(V)=-E\left[\nabla_{\theta}M(\theta)|V \right]E\left[M^2(\theta)|V \right]^{-1}.
\end{align*}
Due to H{\'a}jek's representation theorem  \citep{hajek1970characterization}, the most efficient regular estimator is asymptotically linear and so the existence condition in lemma S1 holds when we consider only RAL estimators. 
\end{proof}

\subsection*{Additional simulation results}

\begin{figure}[H]
   \centering
    \caption{{Boxplots of inverse probability weighted (IPW),  imputation-based (IMP) and doubly-robust (DR) estimators of the regression coefficient $\beta_0$, whose true value of 0.5 is marked by the horizontal line, when $\alpha_3=2$.}}
       \includegraphics[page=1,width=0.8\textwidth]{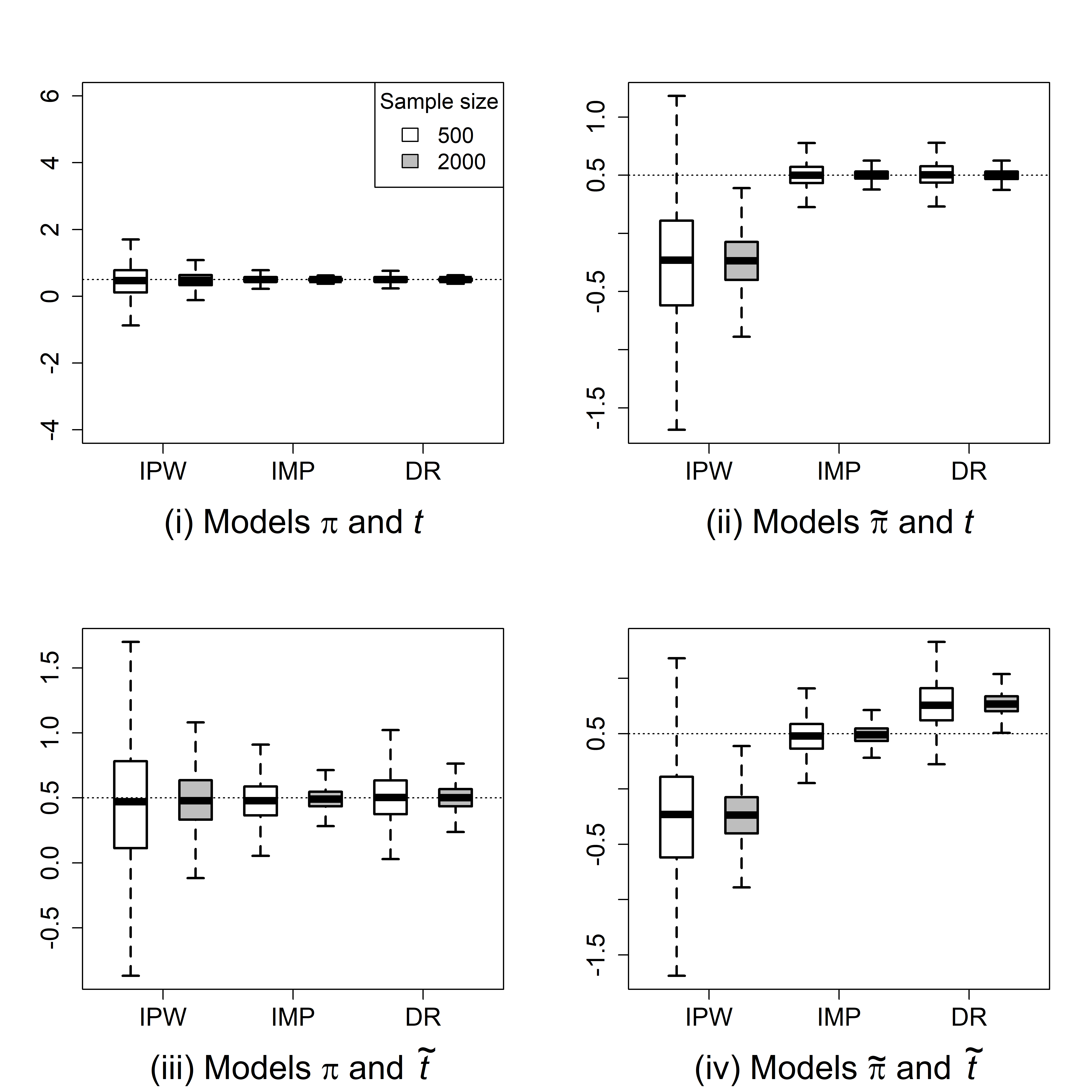} 
        \label{fig:1}
\end{figure}

\begin{figure}[H]
   \centering
    \caption{{Boxplots of inverse probability weighted (IPW),  imputation-based (IMP) and doubly-robust (DR) estimators of the regression coefficient $\beta_1$, whose true value of -0.5 is marked by the horizontal line, when $\alpha_3=2$.}}
       \includegraphics[page=1,width=0.8\textwidth]{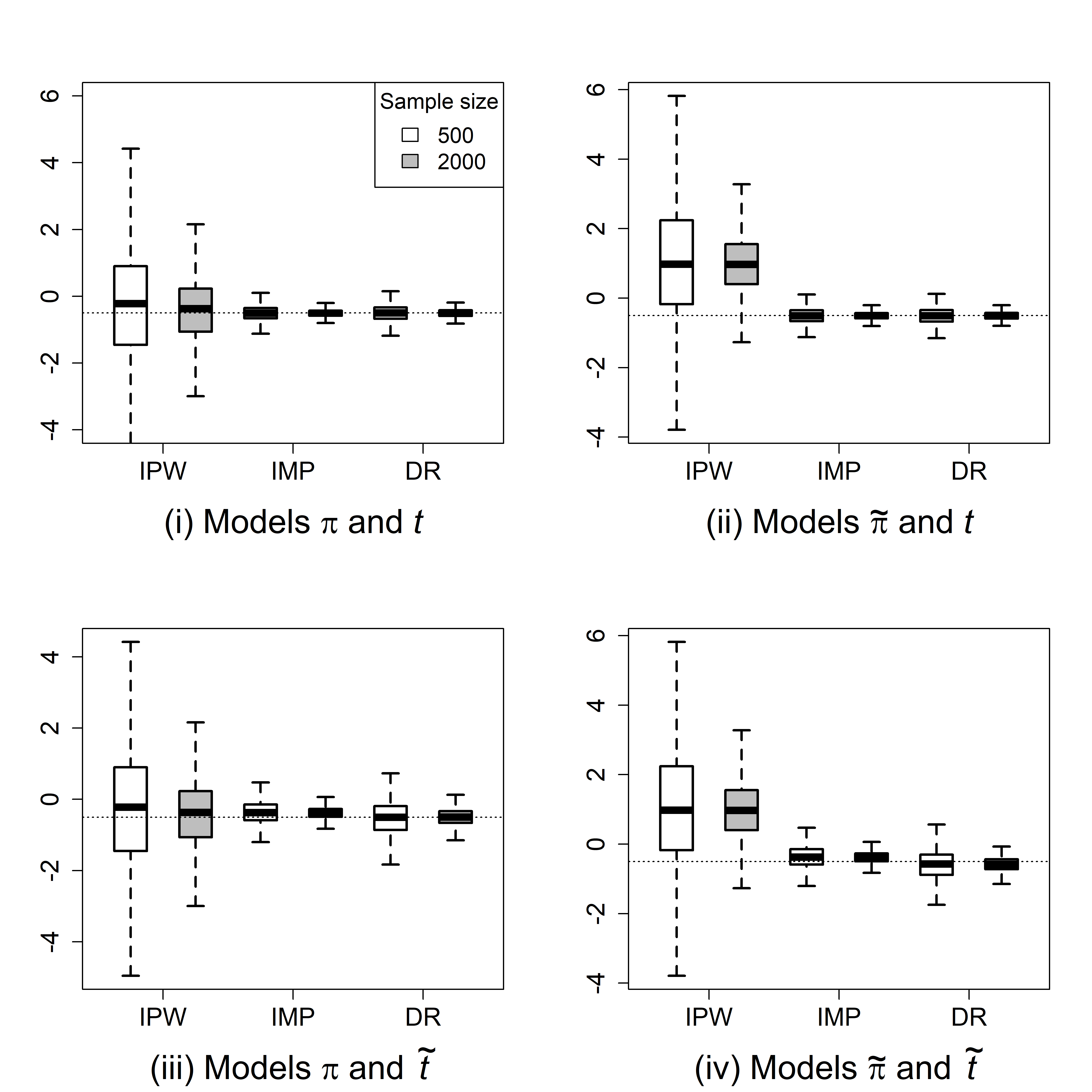} 
        \label{fig:1}
\end{figure}

\begin{figure}[H]
   \centering
    \caption{{Boxplots of inverse probability weighted (IPW),  imputation-based (IMP) and doubly-robust (DR) estimators of the regression coefficient $\beta_2$, whose true value of 1.0 is marked by the horizontal line, when $\alpha_3=2$.}}
       \includegraphics[page=1,width=0.8\textwidth]{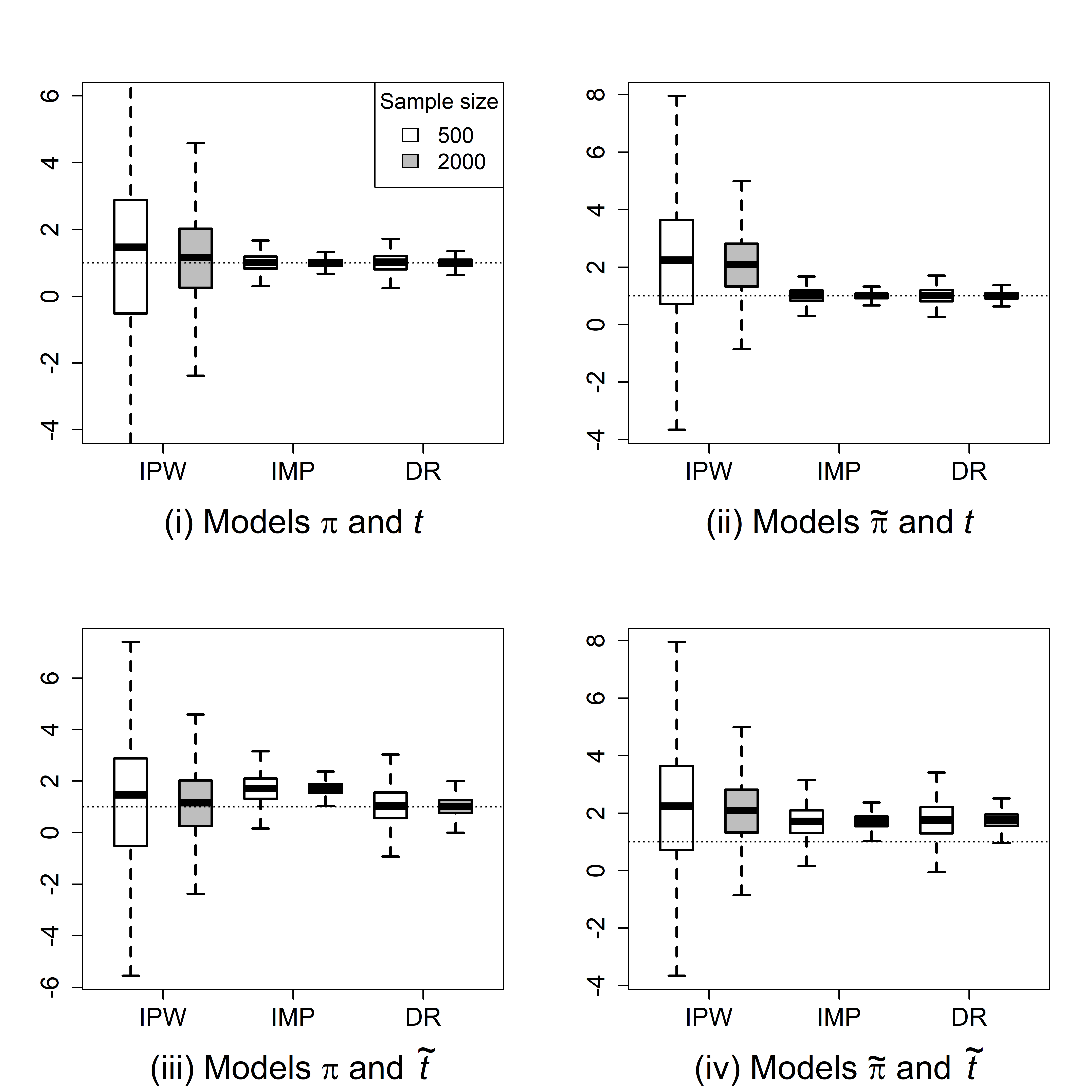} 
        \label{fig:1}
\end{figure}

\begin{figure}[H]
   \centering
    \caption{{Boxplots of inverse probability weighted (IPW),  imputation-based (IMP) and doubly-robust (DR) estimators of the regression coefficient $\beta_0$, whose true value of 0.5 is marked by the horizontal line, when $\alpha_3=0.5$.}}
       \includegraphics[page=1,width=0.8\textwidth]{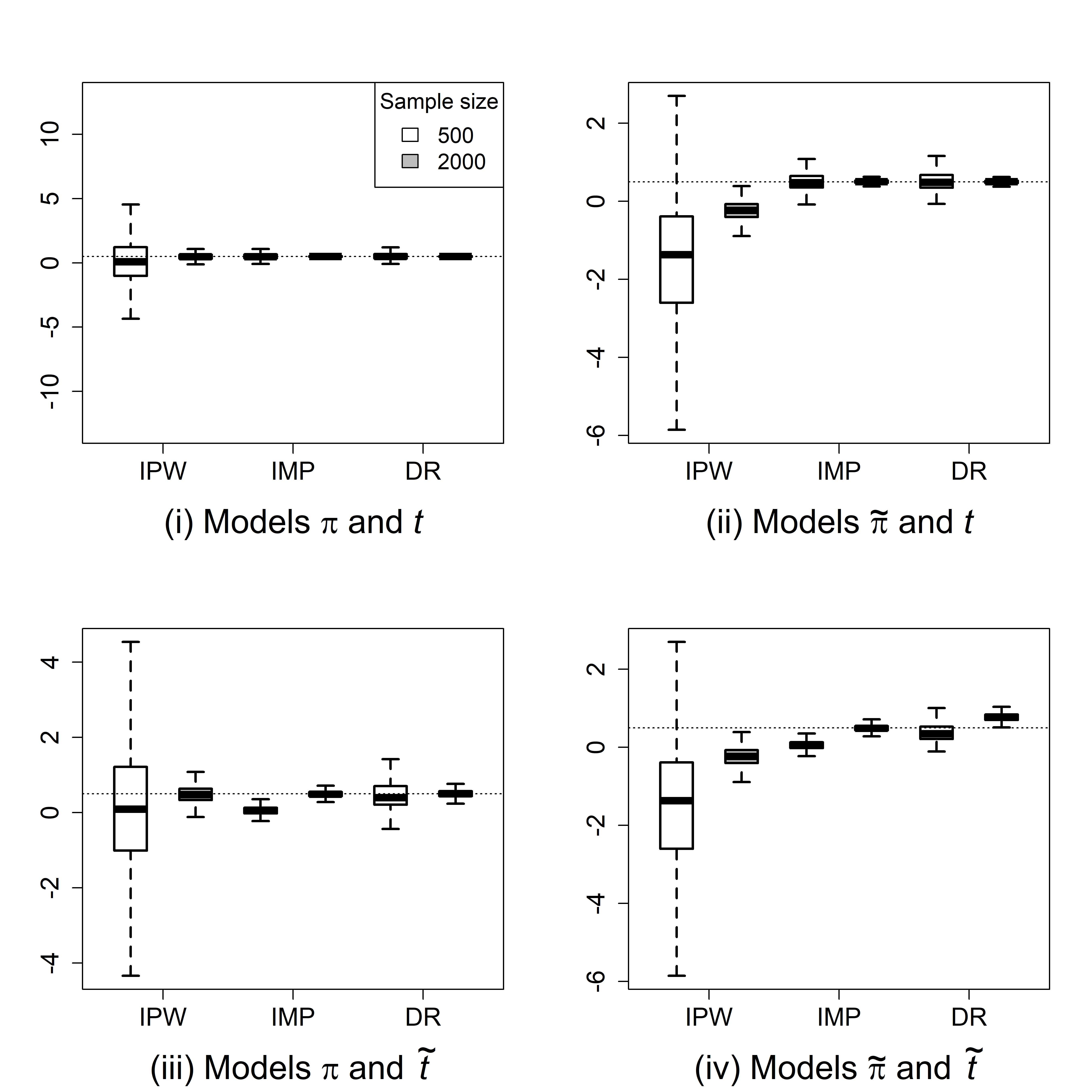} 
        \label{fig:2}
\end{figure}

\begin{figure}[H]
   \centering
    \caption{{Boxplots of inverse probability weighted (IPW),  imputation-based (IMP) and doubly-robust (DR) estimators of the regression coefficient $\beta_1$, whose true value of -0.5 is marked by the horizontal line, when $\alpha_3=0.5$.}}
       \includegraphics[page=1,width=0.8\textwidth]{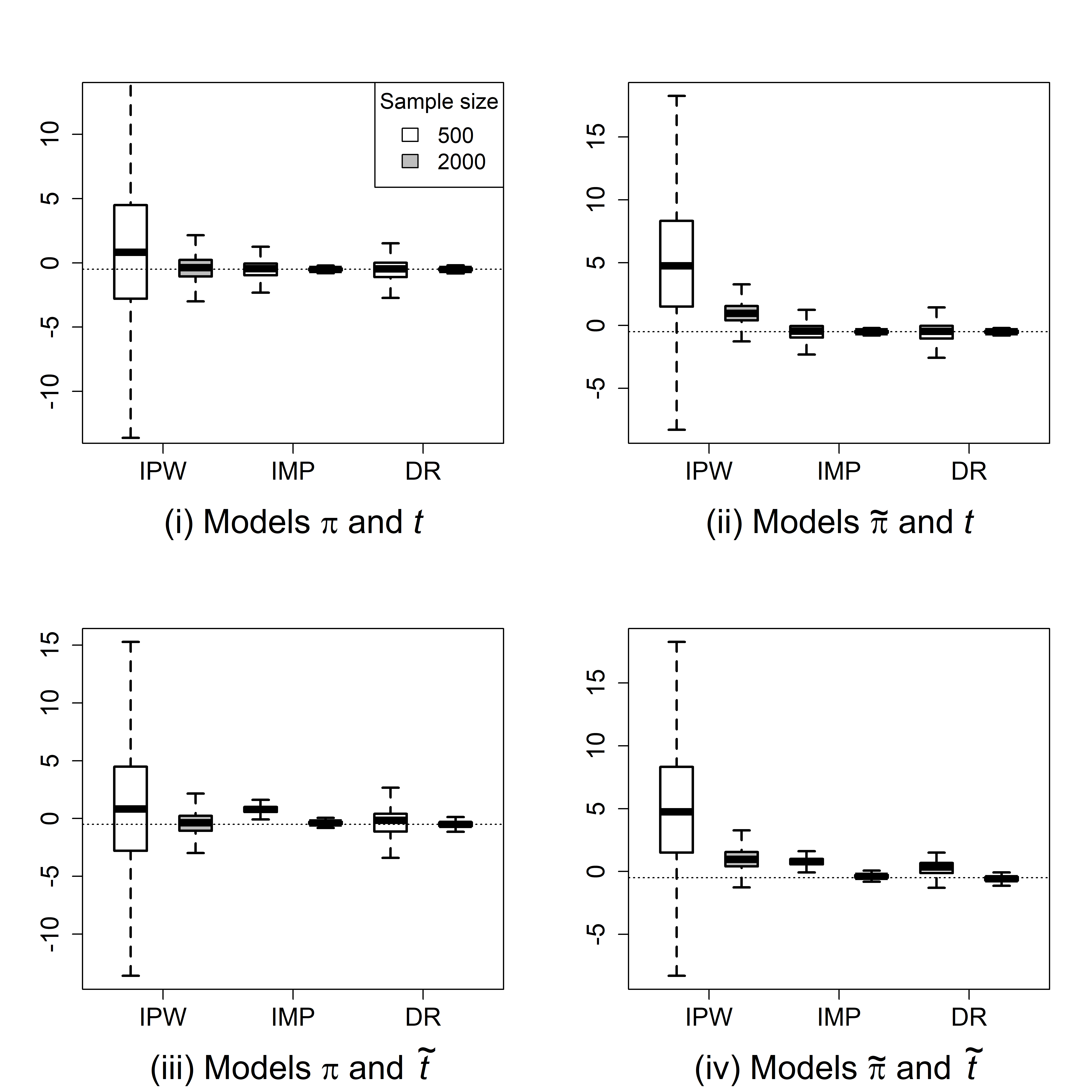} 
        \label{fig:2}
\end{figure}

\begin{figure}[H]
   \centering
    \caption{{Boxplots of inverse probability weighted (IPW),  imputation-based (IMP) and doubly-robust (DR) estimators of the regression coefficient $\beta_2$, whose true value of 1.0 is marked by the horizontal line, when $\alpha_3=0.5$.}}
       \includegraphics[page=1,width=0.8\textwidth]{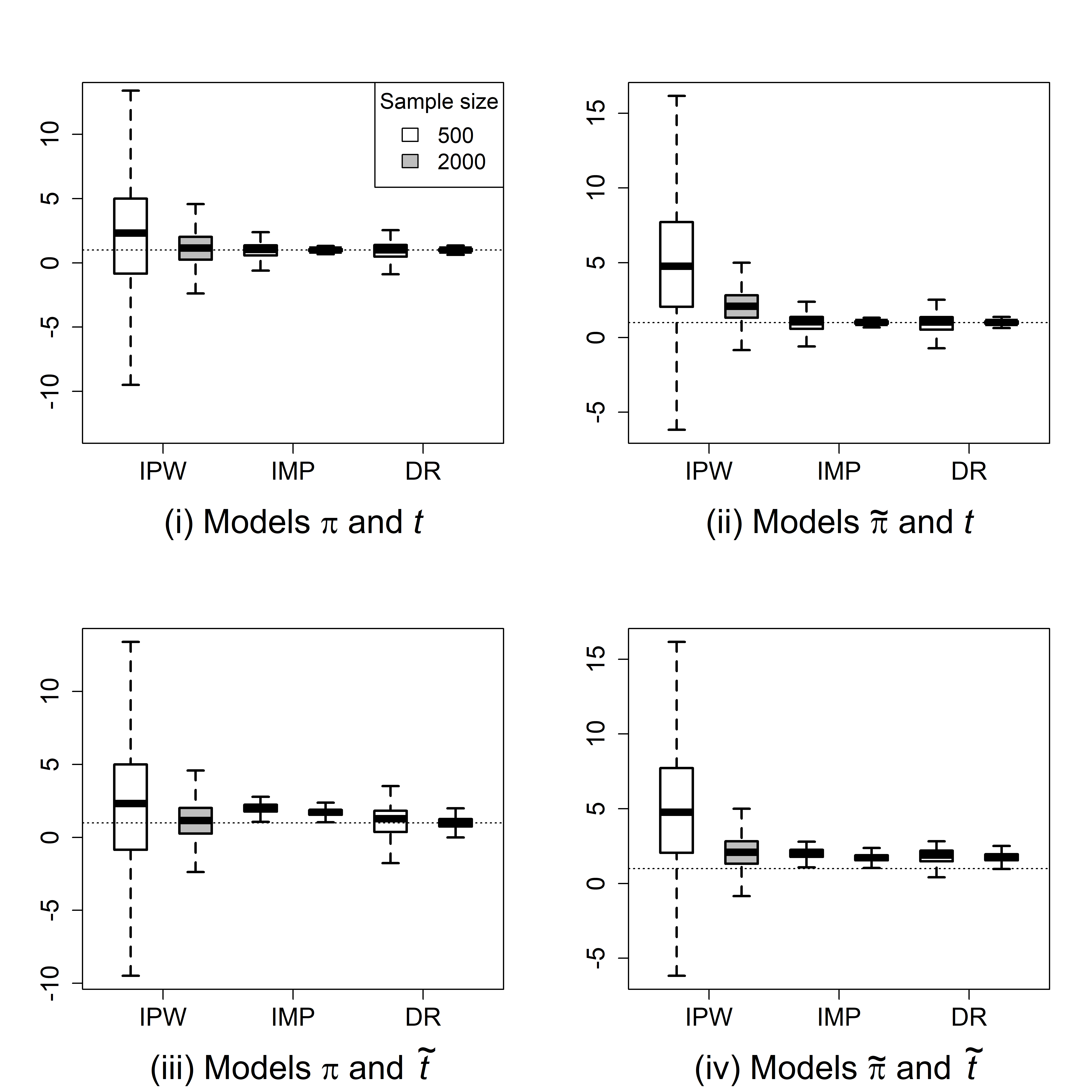} 
        \label{fig:2}
\end{figure}

\end{document}